\documentclass[fleqn,twoside]{article}
\usepackage{espcrc2}

\usepackage{graphicx}
\usepackage[figuresright]{rotating}

\newcommand{\AmS}{{\protect\the\textfont2
  A\kern-.1667em\lower.5ex\hbox{M}\kern-.125emS}}

\title{ The Status of the ANTARES experiment}

\author{E. V. Korolkova\address[MCSD]{Department 
        of Physics and Astronomy, The University of Sheffield, 
	Sheffield S3 7RH, UK}
        \thanks{e-mail: e.korolkova@sheffield.ac.uk}
	for the ANTARES Collaboration}

\begin{document}

\begin{abstract}
ANTARES is a neutrino telescope designed to search for high-energy neutrinos 
from astrophysical sources such as quasars, gamma-ray bursters, microquasars, 
supernova remnants and AGN. The objectives also include the indirect search for 
WIMPs, primary candidates for non-baryonic dark matter, by looking for 
neutrinos from neutralino annihilations in the centres of the Sun, Earth and
Galaxy. The array of 12 lines with 900 photomultiplier tubes will be deployed 
by 2007 at a depth of about 2500 m in the Mediterranean sea near Toulon (France),
40 km off the coast. It will detect the Cherenkov light emitted in sea water by 
muons produced via charged-current interactions of neutrinos with surrounding
matter. A prototype line and an instrumentation line for monitoring 
environmental parameters have been successfully deployed and connected to the 
electro-optical cable, which transmitted the data to the shore station. The 
current status of the project is presented.
\vspace{1pc}
\end{abstract}

\maketitle

\section{INTRODUCTION}

The acceleration of cosmic rays up to very high energies  is one of 
unsolved problems of modern physics. The observed cosmic-ray particles with 
energies up to $10^{20}$ eV demonstrate the existence of 
sources capable of accelerating protons to such energies. Active galactic nuclei, 
supernova remnants, microquasars and gamma ray bursts are considered as candidate 
sources. High-energy particles can be produced in hadronic and 
electromagnetic processes. 
High-energy gamma rays from many point-like  sources have been observed 
by ground-based telescopes. However, observation of gammas cannot reveal the 
hadronic nature of the acceleration process inside the source. Moreover, high-energy 
gammas interact with microwave and infrared cosmic background and cannot 
reach the Earth from large distances.
 
 Neutrinos can be produced in $pp$ or $p\gamma$ interactions of accelerated 
protons (or heavier nuclei) with matter or photon field via the decay of charged 
pions (and possibly kaons).  As neutrinos are  electrically neutral and weakly 
interacting particles, they  can escape from the dense cores of potential sources 
and travel enormous distances without being absorbed, scattered  or deflected by 
magnetic fields, delivering  information  on the processes of particle 
acceleration directly from the sources.
  
However, due to the
extremely low neutrino cross sections with respect to the fluxes from
potential sources, predicted by various models, neutrino detection
requires the instrumentation of large target masses, 
suggesting the use of naturally abundant detection materials, 
such as water or ice. A high-energy neutrino interacts in rock, ice  or  water 
and produces a muon, which  emits Cherenkov light while propagating in water 
or ice. The Cherenkov light 
can be detected by an array of optical sensors, called neutrino telescopes. Two 
Cherenkov neutrino telescopes AMANDA and Baikal  \cite{ama} are already taking data and 
producing interesting results, others, such as  NESTOR, NEMO  and ANTARES   are 
at the construction or evaluation stage. 
IceCube and KM3NET are cubic kilometre scale projects.

\section{DETECTOR DESIGN}

The ANTARES project (Astronomy with a Neutrino Telescope and Abyss environmental 
RESearch)  \cite{ant} has been started in 1996. The ANTARES Collaboration aims to build 
a deep underwater neutrino telescope in the Mediterranean Sea.
The selected site is located at $42^\circ 50'N, 6^\circ 10'E$, about 40 
km off the French coast, at a 
depth of  about 2500 m. The site combines the advantage of large depth with the 
proximity to the coast and  infrastructure (harbours of Toulon and La Seyne). The detector
location provides a 3.5 $\pi$ sr coverage of the sky and allows the observation of the 
Galactic  Centre  for 67\% of the time.   

\vspace{-0.4cm}
\begin{figure}[htb]
\includegraphics[width=7.5cm]{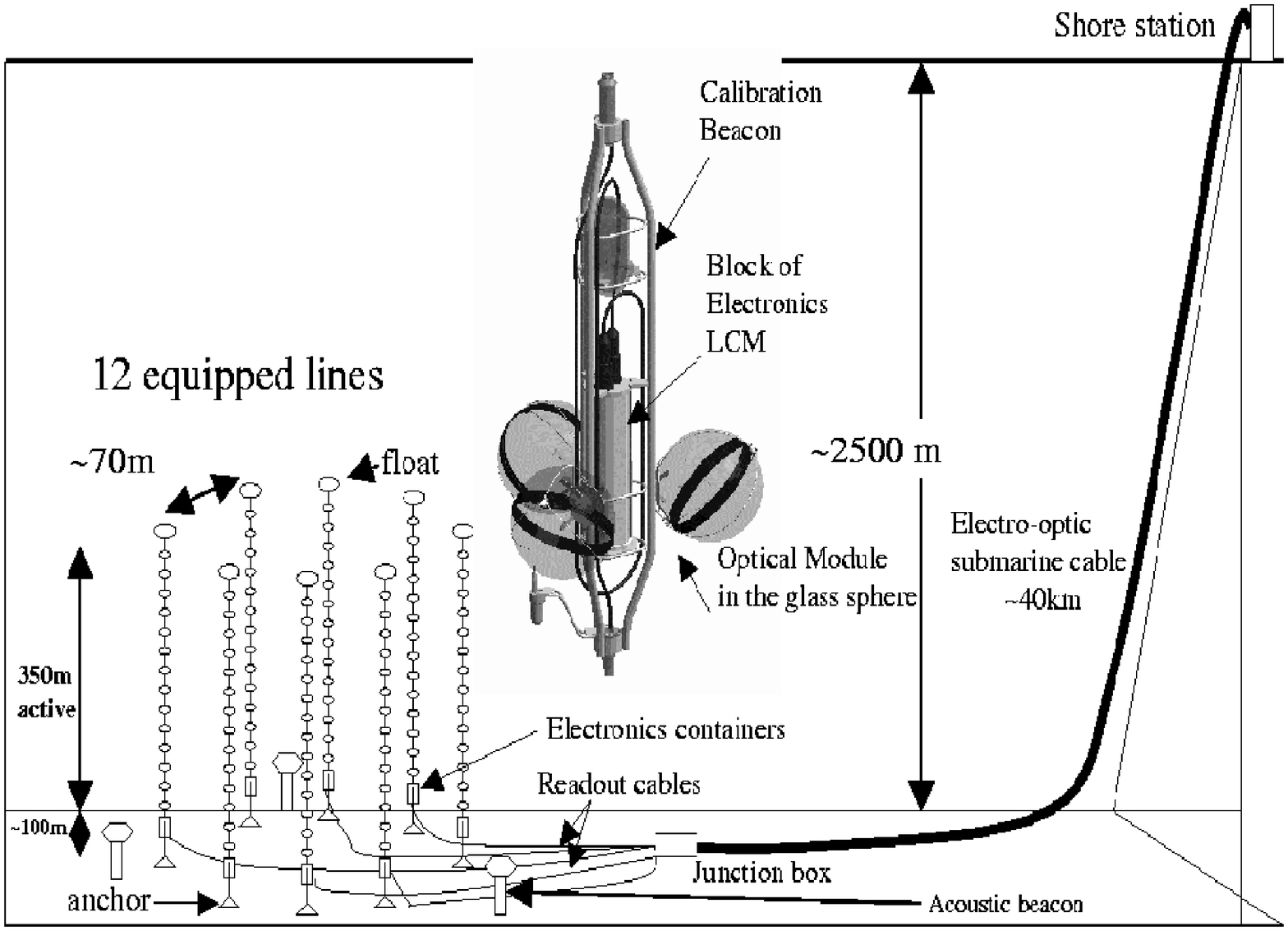}
\vspace{-0.6cm}
\caption{Schematic view of the ANTARES detector.}
\label{fig:det}
\end{figure}
\vspace{-0.4cm}

The choices of the geometrical parameters of the instrumented lines have been guided by the 
results of  measurements performed on water transparency, sedimentation, water 
current and optical background. The absorption length is about 60 m  at 470 nm and mainly
limits the size of the instrumented region and the photomultiplier tube (PMT) spacing. 
The effective scattering length is more than 200 m.  
The optical background is due to  $\beta$ decays of  $^{40}$K and a continuous 
bioluminescence rate of about 60 kHz per PMT which is sporadically increased by short 
bioluminescence
bursts up to MHz. The dead-time due to these processes is a few percent   per PMT randomly distributed
over the detector, however  the requirement for coincidences significantly decreases 
the dead-time for the whole detector.  The average loss in the light
collection at the  optical module due to bio-fouling and sedimentation is less than 2\% in the
horizontal direction
after one year of deployment and it tends to saturate with time \cite{opt}.
 
The ANTARES detector will consist of 12 vertical lines arranged in an octagonal 
geometry (see Figure \ref{fig:det}). Neighbouring lines are separated by about 70 m. 
The lines are anchored 
to the sea bed and kept vertical by buoys attached to the upper ends. 
They float in the sea current and the positions of active detector elements are 
permanently monitored by an acoustic calibration system.
All lines are connected to the junction 
box, which links the detector to the shore station via a 40 km electro-optical 
cable, distributes the electric power and control signals, and sends data to 
the shore station.

Each line is equipped with 75 optical modules (OMs) \cite{nim} 
arranged in triplets (storeys); there are 25 storeys per line. The distance 
between two adjacent storeys is 14.5 m. The storeys are 
interconnected with an electro-optical mechanical cable supplying the electric 
power and the control signals, and transferring the data to the bottom of the 
line. The anchor is placed 100 m below the lower storey. The total height of 
such a line is about 450 m. The detector will be complemented by a special 
instrumentation line equipped with devices 
for monitoring the environmental parameters and tools used for studies in 
oceanography, marine biology and seismology. 

The ANTARES OM is a pressure resistant glass sphere which contains a
hemi-spherical PMT with a photocathode of 10-inch diameter, 
its base and a pulsed LED for timing monitoring. The PMT looks downwards at $45^\circ$ 
to the vertical to avoid sedimentation. 
The PMT is surrounded by a  $\mu$-metal cage which acts as a shield  
against the Earth's magnetic field. A cylindrical titanium container  houses the local electronics.
Some storeys contain  supplementary  calibration equipment like acoustic
and optical beacons. 

The PMTs detect photons with a quantum efficiency above 20\% for the relevant 
wavelengths between 330 nm and 460 nm. The signals of each PMT are read out by two ASICs
(two chips allow reduction of the overall dead time). For simple pulses the charge and
arrival time are digitised and stored for transfer to the shore station. For more
complex pulses the pulse shape is digitised with a 1 GHz frequency. The time stamps are
synchronised by a clock signal which is sent in regular intervals from the shore to  
all electronics cards. 
All signals above the threshold 
(corresponding to about 0.3 photoelectrons) are sent to the shore station where a PC farm 
performs data filtering (the data volume is reduced by a factor of 100), 
selection of event candidates, event building etc. 
These shore computers also facilitate the remote control and the monitoring of the junction box 
and of the lines.

\vspace{-0.5cm}
\begin{figure}[htb]
\includegraphics[width=7.5cm]{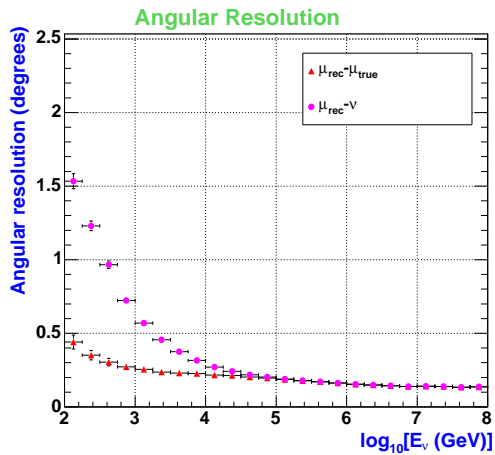}
\vspace{-0.8cm}
\caption{Median angle between the reconstructed muon
and the simulated muon (triangles) or the simulated neutrino (circles) versus 
neutrino energy.}
\label{fig:angle}
\end{figure}
\vspace{-0.5cm}

For a good track reconstruction two main calibrations must be performed accurately. 
The time calibration is
critical for the ANTARES experiment. The time resolution  of the signal pulses will be 
limited by the transition time spread of the 
PMTs ($\sigma$ is about 1.3 nsec).  
Each line is equipped with  
optical beacons (OB) for timing calibration: a laser beacon located at 
the bottom of the line and  LED beacons placed along each detector line. The 
time calibration system aims to achieve a relative precision better than 0.5 ns which 
allows the reconstruction of muon tracks with angular resolution better than $0.3^\circ$
 for energies higher than 1 TeV and absolute time precision of about 1 msec.  

The positional calibration is achieved using a system of acoustic beacons 
placed on and around ANTARES and acoustic detectors (hydrophones) located on 
each storey to measure the position of each OM along the line.  This system 
gives a relative measurement of position of each element of the 
detector with an accuracy of a few centimetres. The absolute calibration comes from the GPS 
measurements of the line positions during  the deployment and  is about 1 metre. 

\section{THE  DETECTOR PERFORMANCE}

 Most studies in the ANTARES programme are concentrated on the charged current interactions 
of muon neutrinos. The detector will operate by detecting the intensity and the 
arrival time of the Cherenkov
light emitted by relativistic charged particles  produced by neutrino interactions. 
The PMT signals are used to reconstruct the muon track. In ANTARES, several muon 
reconstruction algorithms have been developed. They use the direct 
Cherenkov hits but take also into account secondary effects like diffusion, 
dispersion and electromagnetic showers which accompany high energy muons.  
The muon trajectory can be determined from the knowledge of the
arrival times of photons recorded by the PMTs and of their positions. The
pointing accuracy will be better than $0.3^\circ$ for energies above 1 TeV.  
Figure  \ref{fig:angle}  shows the median angle  between the 
reconstructed muon and the simulated muon (triangles) or neutrino (circles) 
versus the neutrino energy. 	
Below 1 TeV the median angle between the muon and the parent neutrino is dominated by 
the kinematics of the interaction, while at larger energies it is limited by the 
intrinsic angular resolution (PMT transit time spread and light scattering 
in water). The limiting values are about $0.15^\circ$.

Different energy reconstruction algorithms  have been developed for the 
ANTARES experiment \cite{icrc_r}. The muon energy can be estimated using the methods 
based on the knowledge of the features of muon energy losses: the increase of the amount 
of emitted light due to muon catastrophic energy losses above 1 TeV. However, the 
measurement is compromised by the fact that these radiative processes are stochastic, 
the neutrino interaction point is invisible in most cases and only a short fraction of 
the muon track is seen in the detector. Since the energy loss of muons at high 
energies has large fluctuations, a parameter used is the logarithmic energy resolution,
$log(\sigma_E/E)$. The simulations show that this resolution is 0.2--0.4 for muons with energy
above 1 TeV. At lower energies, below 100 GeV, the muon energy can be measured from the muon range 
in sea water.  

\vspace{-0.5cm}
\begin{figure}[htb]
\includegraphics[width=7.5cm]{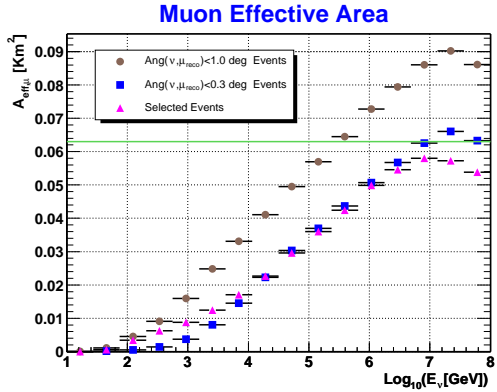}
\vspace{-0.8cm}
\caption{Muon effective area versus neutrino energy 
for  selected events (triangles) and for  
reconstructed events with the requirement that the angle between 
the simulated neutrino and the reconstructed muons is less than $1^\circ$ (circles) and  
$0.3^\circ$ (squares).}
\label{fig:amu}
\end{figure}
\vspace{-0.5cm}

The effective area is one of the important parameters which characterise the performance of 
the detector. The muon 
effective area is the ratio of the rate of the selected events to the flux of incident 
muons and depends on selection criteria used for any specific analysis.   The 
effective surface area of ANTARES depends on the muon (neutrino) energy, the 
efficiency of reconstruction and selection cuts. Figure \ref{fig:amu} shows the muon
effective area as a function of neutrino energy for isotropically simulated events after 
a selection based on reconstruction quality  
cuts (triangles) and just requiring an 
angle between the reconstructed muon and the simulated neutrino direction lower than $1^\circ$ 
(circles) and $0.3^\circ$ (squares). 

For point-like source searches,  when a signal should be found in 
the whole sky and not necessarily associated with any known object, strict 
selection criteria are needed to keep only well reconstructed events. This 
ensures a good pointing accuracy and a good rejection of the background of 
atmospheric muons.  Other searches, for instance for bursting known sources, such 
as gamma-ray bursts, which are almost background free on time scales of hundreds 
of seconds, can be done by replacing stringent cuts on the quality of the 
reconstruction by an angular cut around the supposed known source. In this case 
the effective area calculation is performed by requiring that the angle between 
the direction of the reconstructed muon and the neutrino direction is less than 
a defined angular cut.
For well reconstructed muon tracks (accuracy better than $1^\circ$) the effective 
surface area increases from about $10^{-3}$ km$^2$ at neutrino energy  0.1 TeV to more 
than 0.06-0.07 km$^2$ at neutrino energy higher than 100 TeV. The angular distribution 
is averaged over 
the upward going hemisphere. The increase of effective area with energy is explained 
by the increasing muon range and increasing light output of muons due to radiative 
processes.  

\vspace{-0.5cm}
\begin{figure}[htb]
\includegraphics[width=7.5cm]{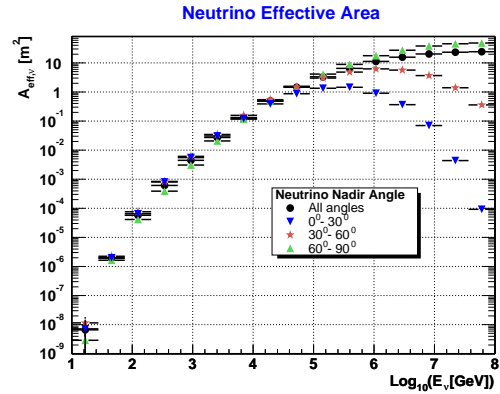}
\vspace{-0.8cm}
\caption{Neutrino effective area as a function of the neutrino energy in ranges of 
neutrino nadir angle.}
\label{fig:anu}
\end{figure}
\vspace{-0.5cm} 

Figure \ref{fig:anu} 
gives the effective area for neutrinos as a function of the neutrino energy
for three different nadir angles and for the averaged over nadir angle area. The neutrino 
effective area is calculated taking into account the detection efficiency, 
the probability of interaction,  
the probability that neutrino survives its journey through the Earth and the energy loss
of muons. Small  neutrino cross sections change the scale of the effective area  from 
squared kilometres to squared metres; 
the opacity of the Earth limits the effective area to values below 30 m$^2$. The 
effect of neutrino absorption is visible since the area for vertical directions 
is smaller than for horizontal ones.

\vspace{-0.6cm}
\begin{figure}[htb]
\includegraphics[width=7cm]{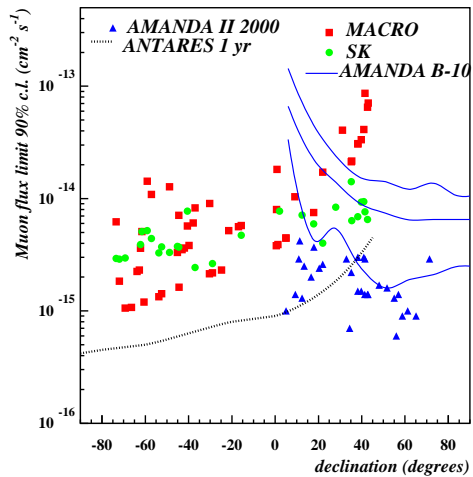}
\vspace{-0.8cm}
\caption{ANTARES sensitivity (after 1 year of operation) to the neutrino-induced flux from 
astrophysical point sources, compared to the upper limits from other experiments (see \cite{point}
for references).}
\label{fig:point}
\end{figure}
\vspace{-0.5cm}

Atmospheric down-going muons and up-going neutrinos constitute the physical 
background for the ANTARES astrophysical programme.  Atmospheric muons, in particular muon bundles, can be 
misreconstructed as up-going neutrino-induced muons. 
This background is currently under study.  Preliminary results show that the 
large depth and sophisticated reconstruction algorithms can help to suppress this 
background sufficiently to be sensitive to astrophysical neutrinos. The signal 
from atmospheric neutrinos is indistinguishable from astrophysical neutrinos.
Using the above performance parameters one can estimate that ANTARES will detect 
about 2500 upward going muon tracks from atmospheric neutrinos per year.
The only difference is the energy spectrum, which is known to be soft for atmospheric
neutrinos (power index of about -3.6), but  is harder for cosmic neutrinos
(power index of about -2).
This background, however, is negligible in a search for point sources of 
high-energy neutrinos, provided the angular resolution of better than $1^\circ$ can be 
achieved in practice.
 
\section{SENSITIVITY TO POINT-LIKE SOURCES, DIFFUSE FLUX AND DARK MATTER}

Detailed simulations have been carried out to assess the physical sensitivity of ANTARES.
Neutrino telescopes may detect astrophysical point-like sources of high energy 
neutrinos as an excess of events above the atmospheric neutrino background.  The 
sensitivity therefore depends on the pointing accuracy of the detector.  Methods 
based on binning or clustering algorithms have been developed, together with a 
method which does not require any binning. The latter is based on a likelihood 
ratio test and uses the information on the neutrino angular resolution as a 
function of energy  \cite{icrc_aart}.

The effective area and angular resolution of the detector determines its 
sensitivity for point-like sources, since the signal--to--noise ratio depends on 
these quantities. The bin size has been optimised in order to obtain the 
discovery potential and exclusion limits for a source with a spectrum proportional 
to $E^{-2}$. It is varied with declination in order to keep constant the average number 
of background events per bin (which is equal to 0.3).
The optimal cluster size for sources with $E^{-2}$ spectrum has been estimated to be 
$1.0^\circ$.
In these methods the significance of a possible excess can be calculated 
analytically from the data itself.  These methods do not rely on the predicted 
spectrum or on the energy reconstruction for further rejection 
of atmospheric neutrinos.

A method not relying on any binning is based on likelihood ratio and 
operates by finding the position  and the flux  of the most likely source 
candidate using the information on the neutrino angular resolution as a function of energy. 
This method gives a 30\% improvement in sensitivity compared to the method based on binning. 
The ANTARES sensitivity to high energy cosmic neutrinos is defined as the upper limit at 90\%  
confidence level (CL) on the muon flux induced by neutrinos, with a typical  $E^{-2}$  energy
spectrum, in the absence of a signal. For declination angles below  $40^\circ$ the
sensitivity, after one year of exposure time, will be  in the range  
$(4-50) \cdot 10^{-16}$ cm$^{-2}$s$^{-1}$ (depending on declination)
for point-like sources, as shown in Figure  \ref{fig:point}.

\vspace{-0.5cm}
\begin{figure}[htb]
\includegraphics[width=7cm]{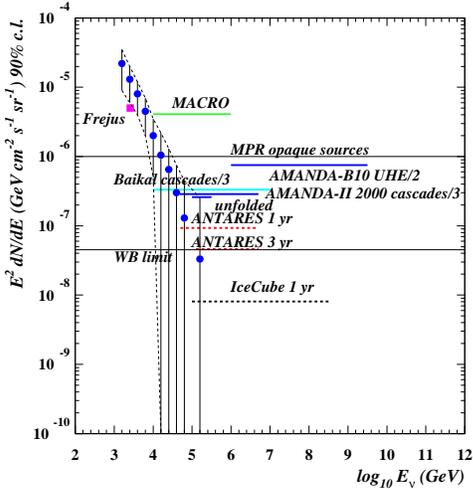}
\vspace{-0.8cm}
\caption{ANTARES sensitivity to diffuse fluxes as a function 
of the neutrino energy. The upper limit that can be set by ANTARES after one year 
and three years of data taking are shown together with  theoretical predictions and 
results of other experiments.  The circles show the atmospheric neutrino spectrum measured by
AMANDA 
(see \cite{dif_fl} for references of the theoretical models and experimental limits).}
\label{fig:diffuse}
\end{figure}
\vspace{-0.5cm}

Diffuse fluxes from astrophysical models are expected to exceed the atmospheric neutrino 
background at energies above 50--100 TeV, so the search for diffuse neutrino fluxes is possible only at 
high energies using the information about the reconstructed muon energy.
After one year of data taking the ANTARES sensitivity to $E^{-2}$ 
diffuse neutrino flux is expected to be
$E^2 d\Phi_\nu/dE_\nu$ $\leq$ $8\cdot 10^{-8}$ GeV cm$^{-2}$s$^{-1}$sr$^{-1}$ and can be
improved to the value of  $3.9\cdot$$10^{-8}$ GeV cm$^{-2}$s$^{-1}$sr$^{-1}$ after 3 years
of data taking that makes the experiment capable of probing the Waxman-Bahcall upper
limit  of $4.5\cdot 10^{-8}$ GeV cm$^{-2}$s$^{-1}$sr$^{-1}$ \cite{WB}.

\vspace{-0.4cm}
\begin{figure}[htb]
\includegraphics[width=7cm]{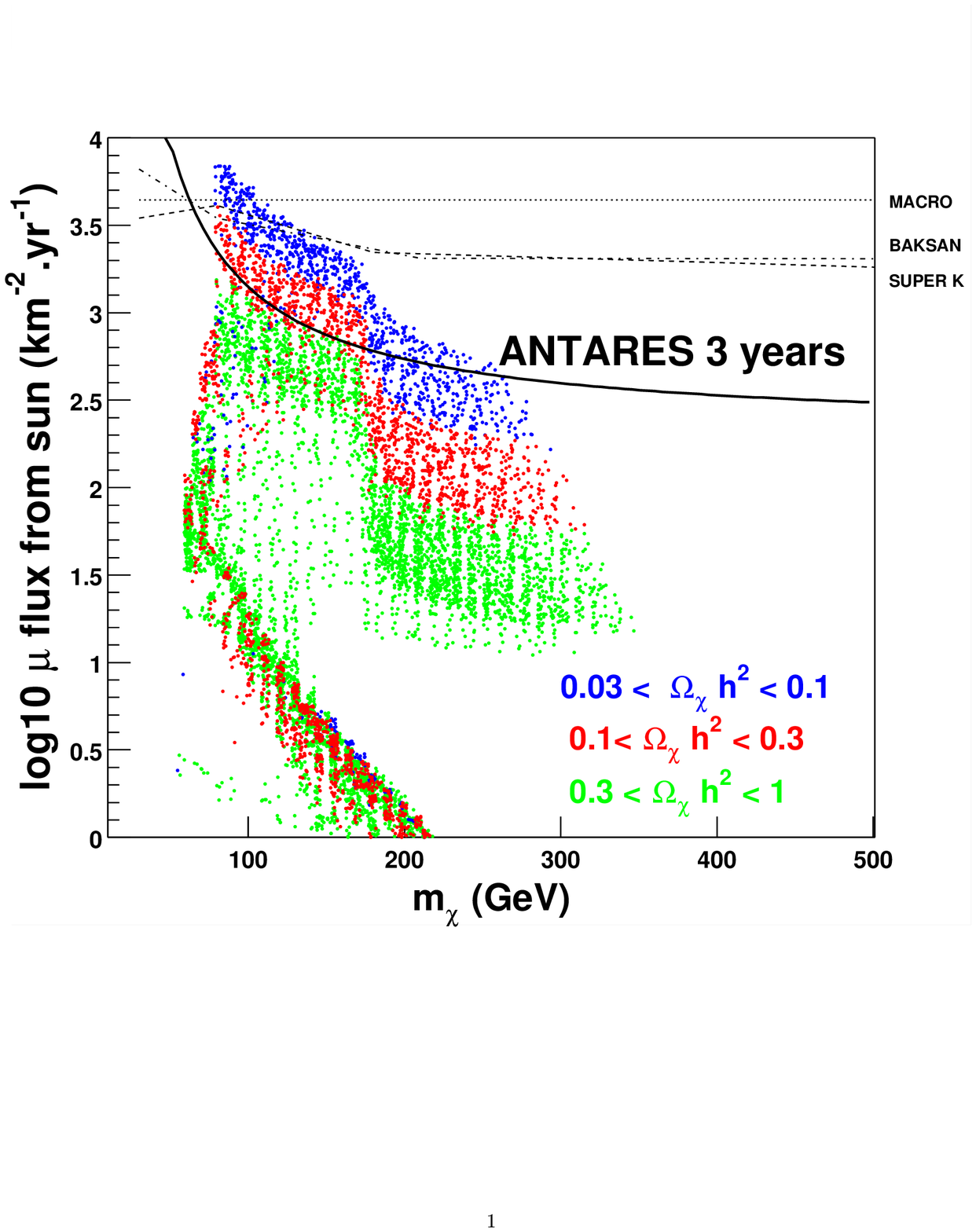}
\vspace{-0.6cm}
\caption{ANTARES sensitivity to the muon flux from neutralino annihilations in the centre 
of the Sun, compared to the upper limits from other experiments and mSUGRA models.}
\label{fig:dark}
\end{figure}
\vspace{-0.4cm}

The search for cold non-baryonic dark matter is an important part of the ANTARES 
scientific program  apart from astronomy. Neutralinos, the best candidates for cold dark matter, 
can be gravitationally captured in the massive astrophysical objects such as the Sun, the 
Earth or the Galactic Centre.
The neutralinos can annihilate producing neutrinos in the decay chain. Expected ANTARES 
sensitivity  to the muon flux  from neutralino annihilation in the centre of the Sun
for the case of a ``hard'' neutrino spectrum (assuming 100\% annihilations to WW)
is shown in Figure \ref{fig:dark} \cite{dm}.

\section{THE CURRENT STATUS OF THE EXPERIMENT}

In 2000 the ANTARES collaboration achieved an important milestone with the deployment and 
operation of a  ``demonstrator line'' equipped with seven PMTs, slow-control devices and an acoustic 
positioning system at a depth of 1100 m. This exercise  confirmed 
the accuracy of relative and absolute positioning to be  about 
5 cm and 1 metre, respectively, verified the validity of the data transfer to the shore station and
allowed the measurements of the angular distribution of atmospheric muons.

Before launching the mass production of the detector elements, to
prove the validity of the final design and to assess the reliability of complex marine operations
for deployment, connection and recovery of the lines in realistic conditions,
it was decided to build and deploy 
two lines: a prototype line consisting of five storeys with 15 OMs (which is a basic building 
block of the detector line) and a mini-instrumentation line (MIL) equipped with  devices for time 
calibration, the positioning system and the instruments for measuring environmental parameters. 
The prototype line has been assembled in summer 2002. In order to verify the functioning of the 
PMTs and DAQ system the line has been tested in the laboratory before deployment. The OMs have
been illuminated by a pulsed laser in the dark room. The most important result is the verification
of the timing accuracy which can be reached by the system. After corrections for 
cable delays the time resolution of the pulses have been confirmed to be 1.3 nsec at the single 
photoelectron level.

In 2001-2003 a series of complex marine operations has been carried out. In October 2001 the 40
km long electro-optical cable which connects the ANTARES site with the shore station
has been deployed. In December 2002 the junction box was connected to  
off-shore cable ending and deployed. Since then the slow control of the junction box has been 
maintained and  shows its very stable behaviour.
Then in December 2002 and February 2003 the prototype and the mini instrumentation lines 
have been deployed and positioned on sea bed within a few metres from their nominal 
positions. In March 2003 a manned submarine Nautile successfully connected both lines to the 
junction box. The data have been taken continuously until the recovery of the prototype line in 
July 2003 and analysed to study the optical background at the ANTARES site.  

Two problems occurred in the prototype tests.  A water leak developed
in one of the MIL electronic containers due to a faulty supplier specification for  
a connector. This made further operation impossible and the line was recovered  in May 2003. Also,
a defect in the clock signal transmission caused by a broken optical fibre inside the line meant that
data with timing information at nanosecond precision were unavailable. 
 
\vspace{-0.4cm}
\begin{figure}[htb]
\includegraphics[width=7cm]{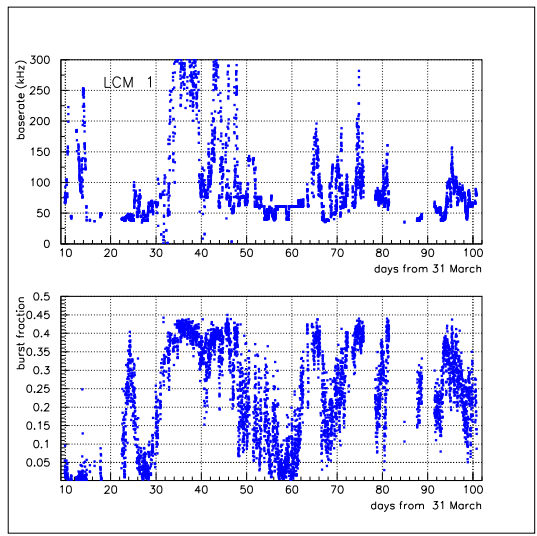}
\vspace{-0.6cm}
\caption{Baseline rate in kHz (top) and burst fraction (bottom) as a function of the
date of measurements. See text for details.}
\label{fig:rates}
\end{figure}
\vspace{-0.4cm}

Nevertheless, during about 100 days of the prototype line operation data were recorded,
both on the functionality of the detector and on environmental conditions. 
In particular,
the rate of signals above threshold was monitored continuously for each OM. It
was found that the rates exhibit strong temporal variations that are attributed
to bioluminescence organisms. A continuous rate, which is caused by $^{40}$K decays and
bioluminescence coming from bacteria, slowly  varying between about 50 kHz
and 250 kHz  per OM, is accompanied by short light bursts of several hundred kHz lasting
from seconds to minutes which are possibly caused by bio-luminescence coming from larger animals.
The top plot of Figure \ref{fig:rates} shows the baseline rate in kHz, which is defined as a 
median of the recorded count rate during a 15 minutes time interval.
The bottom plot shows the burst fraction (defined as the fraction of time when the rate exceeds the 
baseline rate by more than  20\%).
A correlation between bio-luminescence rate and water current velocity has been observed. 
The measured optical background is 50-70\% of 
time below 100 kHz, a rate acceptable for data taking. 
The heading and tilt of the storeys in the prototype line have also been monitored. 
It was found that they move almost
synchronously, i.e. the line behaves as a pseudo-rigid body in  water current.

\section{CONCLUSIONS}
The construction of the ANTARES neutrino telescope has started with the deployment of the main
electro-optical cable and the junction box at the detector site. The experience gained from 
the marine operations validate the detector concept, design and deployment techniques.
The tests with the prototype and mini instrumentation lines have proven to be functional in
real data taking conditions. The data acquisition system has demonstrated its capability 
to cope with high events rate, and the remote control of the lines was fully functional. 
The detected  problems  have been studied, corrected and  will be avoided in future deployments.
Detailed and extensive R\&D studies have been performed.Mass production for the full scale detector have been started. The ANTARES neutrino telescope 
will be fully operational by 2007. Data from the first detector lines are expected in 2005.

\medskip\noindent
{\bf Acknowledgements:} I would like to thank the organisers of the CRIS2004
for their hospitality and some financial support. 
Financial support for the ANTARES project is provided by the following funding agencies:
Commissariat \`a l'Energie Atomique, Centre Nationale de la Recherche Scientifique,
Commision Europ\'eenne (FEDER fund), CPER Alsace: University of Mulhouse, D\'epartement du Var and R\'egion 
Provence Alpes C\^ote d'Azur, City of La Seyne-sur-Mer, France;  
Ministerio de Educaci\'on y Ciencia, Spain (FRA2003-00531);
Istituto Nazionale 
di Fisica Nucleare, Italy; the Foundation for Fundamental Research on Matter FOM and the National 
Scientific Research Organisation NWO, The Netherlands; the Particle Physics and 
Astronomy Research Council, the United Kingdom; Bundesministerium f\"ur Bildung und Forschung (BMBF), Germany.

\end{document}